# The Role of Dark States on Azopyrrole Photoisomerization Reaction by Varying the Coupling Strength of Light-Matter Interaction

Pallavi Garg[‡], Jaibir Singh[‡], Ankit Kumar Gaur[‡], Sugumar Venkataramani[‡], Christian Schäfer[ᴪ]* and Jino George[‡]*

[‡]*Department of Chemical Sciences, Indian Institute of Science Education and Research (IISER), Mohali, Punjab 140306, India.*

[ᴪ]*Condensed Matter and Materials Theory, Department of Physics, Chalmers University of Technology, 412 96 Gothenburg, Sweden.*

**ABSTRACT: Strong light-matter interactions are gaining importance in controlling the photoswitching behavior of molecules exhibiting bistability. Here, we used the strategy of electronic strong coupling of both the *E* and *Z*-isomers of azopyrrole photoswitches to a single cavity mode, thereby tuning the conditions from strong to weak coupling in an on-the-go photoisomerization process. This allows *in situ* monitoring of apparent rates of forward and backward reactions. Very interestingly, the kinetics follow a non-linear trend with a sharp switch in the kinetic slope at the intermediate coupling regime. At this condition, pumping at the upper polaritonic state and the uncoupled population shows an acceleration of the photoisomerization process mediated by the dark state manifold. On the other hand, an opposite effect is observed while exciting the lower polaritonic state. Performing the same experiment in the ultra-strong coupling regime shows no dynamic change over time but further emphasizes the outstanding role of the lower polaritonic states. Our experimental and theoretical findings underline the importance of**

**collective strong coupling in tailoring energy transfer to control photoisomerization dynamically.**

**TOC:**

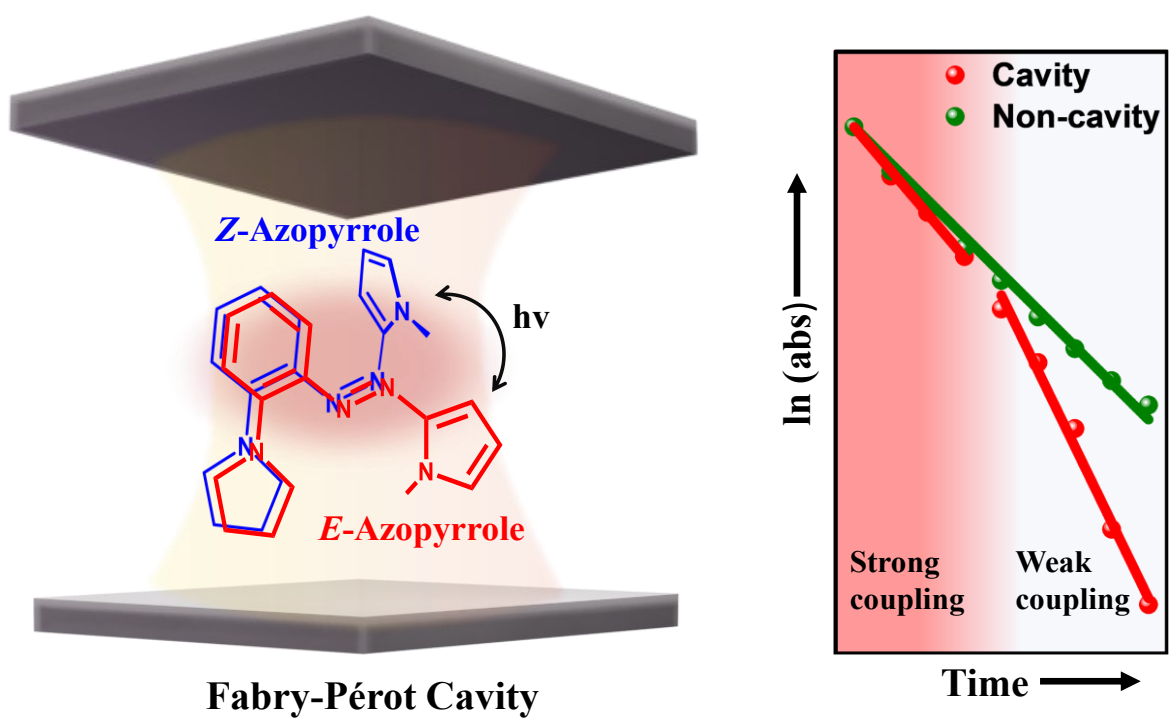

**Introduction**:

Ever since lasers were invented, one of the first thought experiments was to control chemical reactions by external stimuli. Many experiments can be found in the literature, starting from resonant/non-resonant laser excitations, which can selectively target a chemical bond.[1–3] In an unconventional attempt, the last decade witnessed the application of strong light-matter interactions for selective control of chemical reactions.[4,5] This phenomenon replaces external stimuli with vacuum fluctuation (*aka* zero-point energy), leading to the emergence of a new field of research known as *polaritonic chemistry*.[6,7] Strong coupling occurs when a molecular transition interacts resonantly with a cavity photon and overcomes other dissipation rates, forming hybrid light-matter states. The first experimental approach to control a chemical reaction was done on a photoisomerization reaction by electronic strong coupling,[4] which slowed down spiropyran to merocyanine conversion.[5] Recent literature reports suggested that many photoinduced processes can be modified by coupling to the vacuum field.[8–12] At the same time, there are claims showing negligible change in the photoisomerization process if the excitation is done at different conditions.[13] This includes pumping at the polaritonic states versus uncoupled/dark state manifold and also on the choice of non-cavity control experiment used to compare the reaction rates. Excitation done on the active side of a single mirror half-cavity is also found to enhance the reaction rates due to changes in the effective photon flux felt by the molecular switches.[14] These studies motivated us to try a system that has both *E* and *Z* isomers absorbing at the same energy that can be a good test bed for verifying polaritonic effects on photoisomerization processes.

The idea of electronic strong coupling and chemical reaction control triggered an uproar in the research community. There are many theoretical suggestions as to how the polaritonic states can store photochemical energy, thereby slowing down the conversion yield.[15] Standard photoswitches such as azobenzene and stilbene are a few of the molecules tested, and the results indicate a modification of the potential energy surfaces. Ultra-strong coupling (USC) can store the reaction energy for a longer time, thereby almost stopping a photoconversion process.[15] Some of these experiments also propose the photon mixing fraction can tailor the reaction channels, especially the conical intersections, which are important in deciding the energy flow in such systems.[16–18] Exciton-plasmon coupling is also being studied both experimentally and theoretically shown to affect the conversion yield.[19,20] Though these events take place in the excited states, the thermal return channel in such systems is still active and has been found to be almost independent of the excited state reaction dynamics. Further, these experiments pave the way to test the involvement of the ground state in chemical reactions by coupling vibrational transitions.[21] Ground state control of thermal reactions by strong coupling is another attractive challenge for selective control of chemical reactivity.[22,23] Here, the reaction yield is affected by the complex interplay of inter/intramolecular vibrational relaxation channels.[24] The latest understanding of this topic is discussed in some of the review papers.[25–29] Normally, T-type photoswitches exhibit reversibility, and hence, access to the thermal channels can be measured by observing the smooth return of the photoisomers back to their stable form or their native state. Even though the thermal channel can be decoupled with the excited state dynamics, it is more intuitive to test the electronic strong coupling modification of an opposing reaction that can lead to some knowledge of the equilibrium (photostationary state-PSS) in the system.

In the spiropyran-merocyanine experiment, the excitation is done at the isosbestic point in the UV region, and this high-energy pumping is efficient due to the availability of a high transmission window for Ag mirrors.[5] Similar to this experiment, the effect of strong coupling on the photoisomerization kinetics of fulgide is also tested, where the forward rate is found to be accelerated under strong coupling. In the above case, the authors suggest that excited state symmetry is responsible for increased reaction rates.[9] In a different study, photoisomerization quantum yield is calculated for norbornadiene-quadricyclane photoswitches. Here, the photoisomerization quantum yield is similar for coupled and bare molecules when the upper polariton (UP) or the molecular transition is excited. However, the quantum yield conversion drops if the lower polaritonic (LP) state is excited.[8] This intriguing feature suggests a complex interplay of strong coupling that controls the excited state reaction dynamics. A polariton funneling mechanism is also explained in a recent experimental observation.[10] In the current system, we are trying to couple both the *E* and *Z*-isomers of a molecule to a cavity photon and test the photoisomerization process. What if both the isomers are coupled to a cavity? Shall we understand the signature of light-matter interactions by closely observing the system while evolving from strong to weak coupling regime?

**Results**:

Here, we selected (*E*)-1-methyl-2-((2-(pyrrolidin-1-yl)phenyl)diazenyl)-1*H*-pyrrole (hereafter mentioned as azopyrrole) photoswitch - it is an unsymmetrical, push-pull type, and heteroazoarene-based photoswitchable molecular system (**Figure 1a**).[30] This thermodynamically stable *E* isomer can be converted into its metastable *Z* isomer upon photo-irradiation. The metastable *Z* isomer can be converted back to the *E* isomer thermally with a half-life of approximately five minutes in the solution phase. In agreement with previous studies mentioned in the introduction, we found that the strongly coupled states slow down the photoisomerization rate. However, while moving from strong to weak coupling, the isomerization rate shoots up and becomes double at weakly coupled conditions while exciting at the UP and dark state (DS) manifolds. Whereas the situation is exactly the opposite when pumping at the LP state. Interestingly, there is a sudden change around the transition from strong to weak coupling regime. We complement this study with theoretical estimates for the modification of the potential energy surfaces as well as simple models based on state overlap functions, further detailed in the following sections.

Azopyrrole system shows notable advantages over conventional photoswitches. These molecules have a larger separation between π–π* and n–π* transitions, and hence, the mixing of orbitals is less compared to azobenzene. However, the presence of an electron donor pyrrolidine ring at the *ortho* position to the azo unit led to longer wavelength strong absorption in the **n–π*** region due to intense charge transfer in the system. **Figure 1b** shows the absorption spectrum of the *E* and *Z*-isomers. *E*-azopyrrole has two prominent bands, one at 362 nm due to

π–π* transition and a broad n–π* transition with λ$_{max}$ at 490 and 520 nm. We can clearly observe a hump in the broadband absorption spectrum, which can possibly be due to the presence of different conformers. Additionally, the **n-π*** transition is much more intense than π-π*, making it a visible light photoswitch. This is due to better overlapping of the **n** and **π*** orbitals and charge transfer in the system due to the loss of planarity, making the transition intense.[31,32] The extinction coefficient of *E* isomer **n-π*** and **π-π*** transitions are calculated in the toluene solution and found to be 12605 M$^{-1}$cm$^{-1}$ and 7995 M$^{-1}$cm$^{-1}$, respectively.

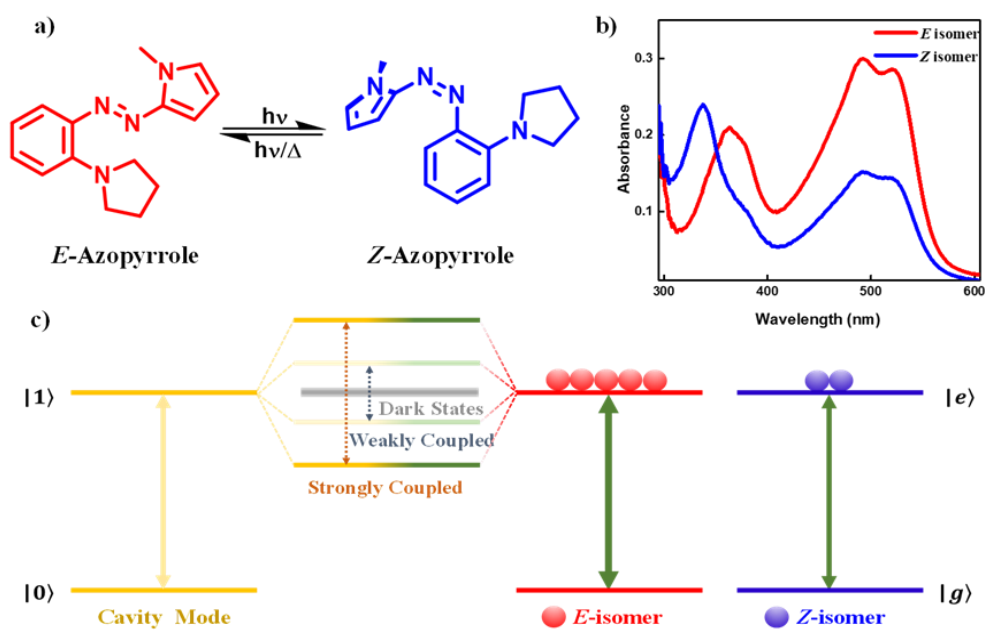

**Figure 1.** a) Molecular structure and photoisomerization reaction of azopyrrole; b) absorption spectra of *E* and *Z*-isomers of azopyrrole derivative in a polymethyl methacrylate (PMMA) matrix; c) Schematic of energy diagram showing the collective coupling of *E* and *Z*-isomers with a cavity mode and the collapse of Rabi splitting energy upon irradiation. Dark states are also shown for clarity. (The relative population difference in *E* and *Z*-isomers is exaggerated).

Interestingly, the **n-π***  transition/CT band for both *E* and *Z* isomers has the same energy but different extinction coefficients; thus, we could couple both the isomers cooperatively to the cavity. The overlapping of the absorption band of the **n-π*** transition makes it an ideal system for testing the effect of strong and weak light-matter coupling on the reaction rates. It is noteworthy to mention that all the strong coupling experiments studied till now involve photoswitches where bond breaking is involved. Whereas, in the current system, there is no bond breaking; instead, the C-N=N-C part of the molecule lies on an inversion/rotation/distortion coordinate that controls the *E*/*Z* photoisomerization. **Figure 1c** depicts the on-the-go change in the Rabi splitting/DS population when it undergoes photoisomerization. Both *E* and *Z*-isomers are cooperatively coupled to the cavity mode due to the fact that they absorb at the same wavelength and give two polaritonic states separated by Rabi splitting energy. Considering the difference in the extinction coefficient between the *E* and the *Z*-isomers, upon photoirradiation, a large decrease in the absorption was observed at the **n-π\*** transition, leading to an effective drop in Rabi splitting. After a certain time interval, the system undergoes transitions from the strong to weak coupling domain, which can be closely monitored, through which its impact on the photoisomerization process under such conditions can be realized.

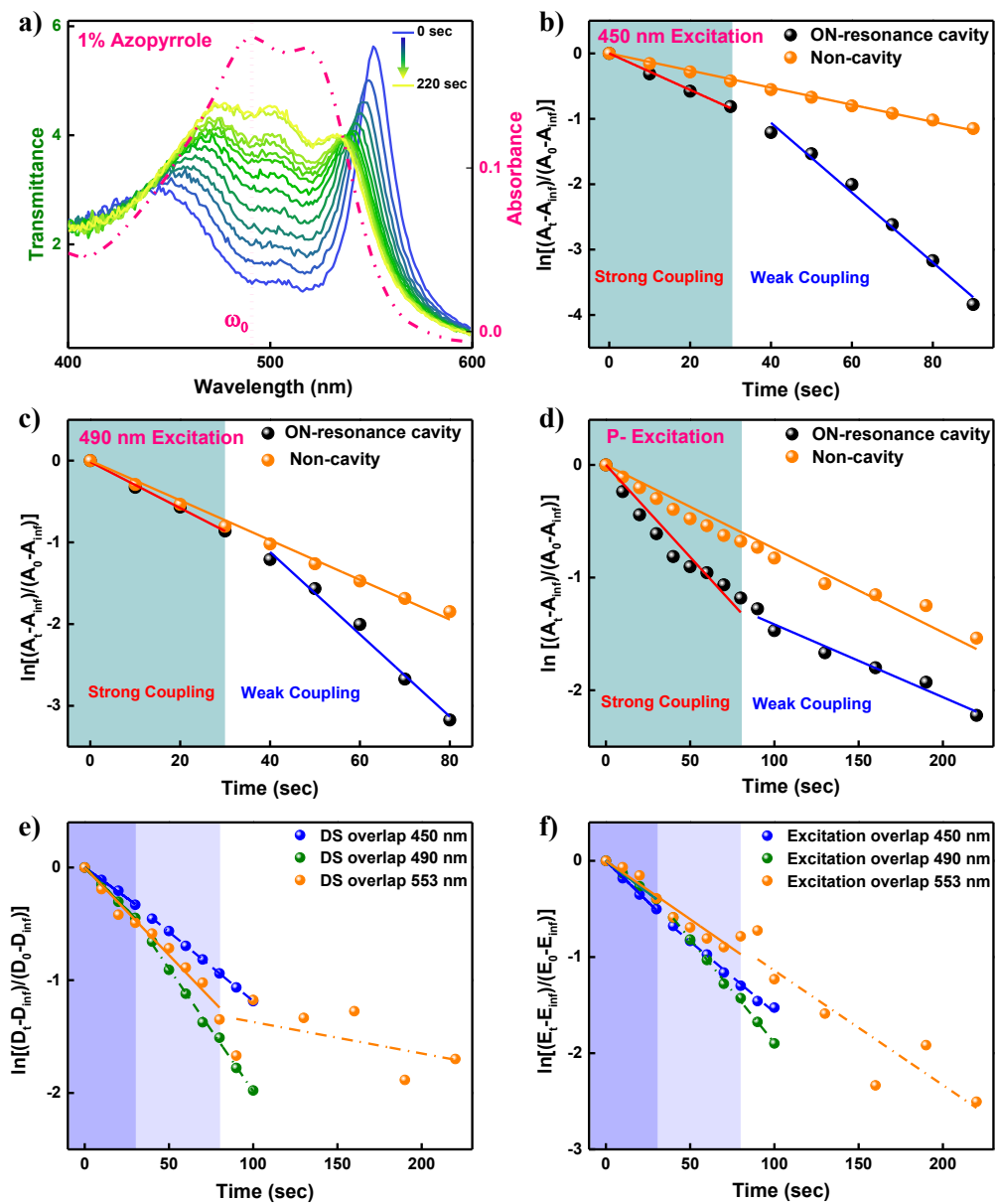

**Figure 2.** *Strong coupling regime*: a) Temporal evolution of *E* to *Z*-isomer conversion of azopyrrole (1 w% in a PMMA matrix) inside the cavity upon photoirradiation with 490 nm LED source; dashed-dot line represents the absorption of *E*-isomer at t=0 (dashed pink line is marked at the absorption maximum). Comparison of photoisomerization reaction kinetics of non-cavity (orange dot) with ON-resonance cavity (black dot) by pumping at b) 450 nm, c) 490 nm, and d) LP (553 nm) excitation energies. Strong (green shade) and weak coupling regimes are demarcated as measured by the absorption linewidth of the molecular band. The theoretical spectral overlap for the time-dependent evolution of polaritonic state e) with the uncoupled molecular line shape (DS overlap), and f) with the light source (excitation overlap) for three different excitations (at 450, 490, and 553 nm). Changes in the DS overlap (e) at 30 seconds (for 450 and 490 nm) and 80 seconds (553 nm) are qualitatively consistent with the experimentally observed rate changes from strong to weak coupling.

After optimizing the spectral features during the photoisomerization of the non-cavity system, we moved into the fabrication of cavity structures. For film preparation, a toluene solution was prepared with 1% azopyrrole and 2% PMMA (w/w) as the final concentration. This freshly prepared solution is spin-coated on an optically clean glass substrate. Further, 30 nm Ag mirrors are sputtered, and the sandwiched configuration forms the FP cavity. The quality factor of the cavity is determined to check the dephasing rate, and it is found to be ~15. Next, we compared the Rabi splitting energy with the FWHM of the absorption spectrum of the molecule to ensure that the system is strongly coupled at *t=0*. At this condition, only the *E*-isomer is present, and the initial Rabi splitting energy is purely dependent on the concentration of this isomer. **Figure 2 (a)** shows the temporal evolution of polaritonic states with 490 nm photoirradiation at a regular time interval of 10 seconds up to 220 seconds. The Rabi splitting energy is with 550 meV (*t=0*) greater than the FWHM of the molecule (450 meV) and the empty cavity mode. To check the kinetics of photoisomerization, both cavity and non-cavity samples were irradiated at absorption maxima of the molecule (490 nm), as well as near the UP (450 nm) and LP positions. As photoisomerization takes place from the *E* to *Z*-isomer, a decrease in the transition probability of the **n–π\*** band takes place, leading to a reduction in the coupling strength. Thus, UP and LP states start moving towards each other, collapsing the Rabi splitting, and pushing the system into the weak coupling regime. **Figure 2 (b-d)** illustrates the reaction kinetics of photoisomerization of azopyrrole inside the cavity and its comparison with the non-cavity. The kinetics is estimated by measuring the absorbance at different time intervals and when the PSS is reached, i.e., $A_\infty$ (Equation 1). The first absorbance value inside the cavity ($A_0$) and the empty cavity mode position ($\omega_c$) are calculated using TMM simulations. Here, we have taken $\omega_c$ as a fixed value during the course of the reaction, assuming that there is no change in the background refractive index ($\Delta n = 0.03$) of the thin film upon photoirradiation. Absorbance values at different time intervals are calculated from their direct relation with Rabi splitting energy as per Equation 2.[8]

$$\ln \frac{(A_0 - A_\infty)}{(A_t - A_\infty)} = -kt \qquad (1)$$

$$c_t = c_{t=0} \left( \frac{\hbar \omega_{R,t}}{\hbar \omega_{R,t=0}} \right)^2 \qquad (2)$$

At 450 nm excitation and under standard conditions (non-cavity), the apparent rate is estimated to be $1.24 \times 10^{-2} \pm 0.10 \times 10^{-2}$ s$^{-1}$ (**Figure 2b**). Within the cavity, we identify two slopes that clearly indicate an increase in the reaction rate when the system transitions from the strong to weak coupling regime. Initially, when azopyrrole is under strong coupling, it follows the first-order kinetics with a rate constant of $2.93 \times 10^{-2} \pm 0.63 \times 10^{-2}$ s$^{-1}$. When the system enters the weak coupling regime, a new first-order curve is observed with an apparent rate of $5.15 \times 10^{-2} \pm 0.77 \times 10^{-2}$ s$^{-1}$. This transition is also indicated by the reduced Rabi splitting of 428 meV at 30 sec, which is now less than the FWHM of the molecule. The same abrupt transition in the isomerization rate is observed when excited at 490 nm (**Figure 2c**). In stark contrast, photoirradiation at the LP position results in the opposite trend, nearly cutting the rate in half at 80 sec (**Figure 2d**). The effect of ESC on the thermal return channels is less prominent, and the subtle changes are due to the differences in PSS in the cavity and non-cavity conditions, which cannot be measured quantitatively.

A key observation is that the sudden change in reaction rate, moving from strong to weak coupling, depends on the excitation energy. Especially puzzling, existing theoretical models based on modified potential energy surfaces would not explain the sudden reduction in rate when exciting the LP state, but would suggest rate enhancement, thus calling the widely accepted explanation in question. One possible alternative mechanism would involve a fast relaxation from the delocalized bright state into the DS manifold with subsequent localization of the excitation on a single molecule that results in the isomerization reaction, as suggested by Dutta A. et al. in a recent article.[12] Following this hypothetical mechanism, the isomerization rate in the cavity is then dominated by the spectral overlap between the time-dependent optical spectrum (excited polaritonic states) and the dark state manifold located at the original maximum of the bare absorption spectrum – essentially enforcing energy conservation for transmission into the DS. We approximate this DS overlap by integrating the overlap of the transmission spectrum of the polaritonic state evolution with the uncoupled molecular line shape of the *E*-isomer absorption spectrum at $t=0$. Further fitting the changes of the overlap integral on a first-order kinetic equation is found to be consistent with the experimentally observed reaction rate. This also reproduces the transition between strong and weak coupling regimes. Here, UP and LP states show the opposite trends, which is exactly reproduced in the DS overlap correlation (**Figure 2e**). Calculating the excitation overlap, i.e., the spectral overlap between the pump source and dynamic absorption spectrum, shows inconsistencies with the observed trend in the experimental reaction rate (**Figure 2f**). This disqualifies such trivial explanations and hints at a mechanism mediated by strong coupling that goes beyond absorption matching.

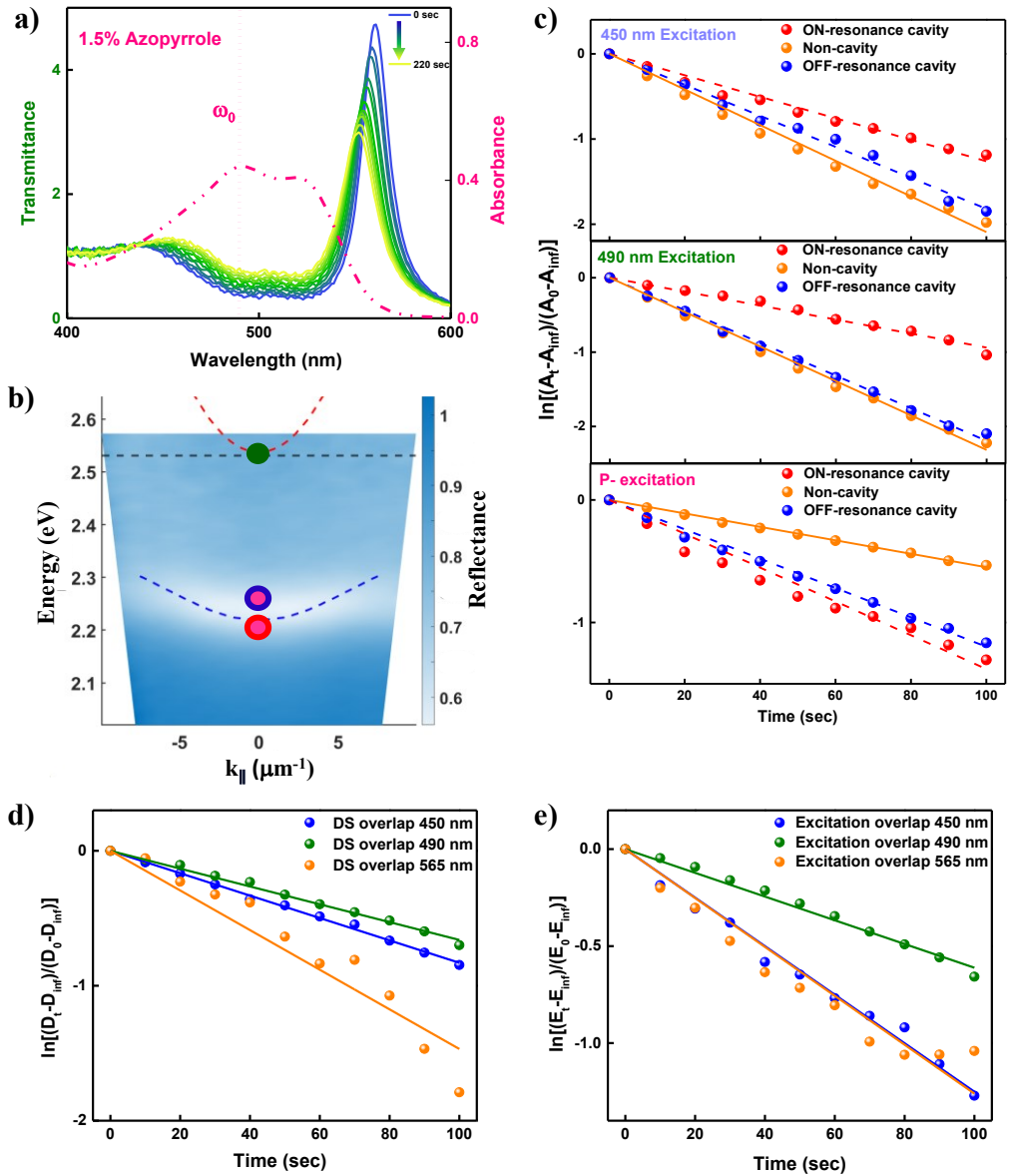

**Figure 3.** *Ultra-strong coupling regime*: a) Temporal evolution spectra of *E* to *Z* conversion of azopyrrole (1.5 w% in a PMMA matrix) inside the cavity upon irradiation with 490 nm LED source (dashed pink line; taken at the absorption maximum). b) Experimentally measured dispersion plot of the ultra-strong coupled system using back-focal microspectroscopy. The black dashed line represents the absorption maximum of the *E* and *Z*-isomers. The dashed line represents the corresponding transfer matrix simulation (TMM) fitting of the empty cavity mode (red) and lower polaritonic state (blue). Excitation positions are represented as green (UP) and pink (LP) (Outline: blue and red corresponding to ON- and OFF-resonance cavity, respectively.) c) Comparison of photoisomerization reaction kinetics of non-cavity (orange dot), ON-resonance (red dot), and OFF-resonance cavities (blue dot) by pumping at 450 nm (first panel), 490 nm (middle panel), and d) LP excitation energies (lower panel). The theoretical spectral overlap for the time-dependent evolution of polaritonic state d) with the uncoupled molecular line shape (DS overlap), and e) with the light source (excitation overlap) for three different excitations (at 450, 490, and 565 nm).

In order to get a clear picture of the role of the DS and its influence on the reaction rate, we further checked the USC regime. Here, we increased the concentration of the azopyrrole molecules in the cavity and non-cavity by preparing 1.5% azopyrrole and 1.5% PMMA (w/w) in toluene. Rabi splitting energy of 626 meV (a change of >20% of the fundamental transition energy of the molecule) was obtained at *t=0*, defacto moving the system into the USC domain. **Figure 3a** shows the evolution of UP and LP states in the USC cavity. Back focal spectroscopy is done for the USC sample, and the LP state is dispersing and is further fitted using TMM simulation. The green, blue, and red dots in **Figure 3b** represent the excitation position at the normal incidence of the cavity. Back focal imaging setup is limited in the spectral window, and hence, the information on the high-energy UP state cannot be obtained. As per the dispersion plot, there is less optically active population at the uncoupled molecular position ($\omega_0$), resulting in a reduced DS overlap. Specifically, the DS overlap at t=0 reduces in USC for LP by 22%, UP by 53%, and 490 nm by 54%. Contrary to the previous case, the system never transitions into the weak coupling domain, giving rise to a single decay rate for the entire measurement time (**Figure 3c**). Even after 100 seconds of photoirradiation, the Rabi splitting is found to be 484 meV, which is greater than the FWHM of the molecule. This indicates that the reaction rate is drastically slowed down under ESC and it is clearly visible from the temporal plot (**Figure 3c**). The reaction rate is calculated using Equations 1 and 2. Exciting UP or 490 nm results in a reduced rate compared to the non-cavity, with a notable resonance dependence. In contrast to the previous excitation energies, the bare spectrum features vanishing oscillator strength above 550 nm, resulting in slow photoisomerization in the non-cavity. ON-resonance (red) and OFF-resonance (blue) cavities show almost similar trends due to the funnelling of low-energy photons more effectively into the photoisomerization process. The resonance dependence is further discussed in the next session. DS overlap (**Figure 3d**) provides again a consistent figure of merit for the experimentally observed reaction rate.

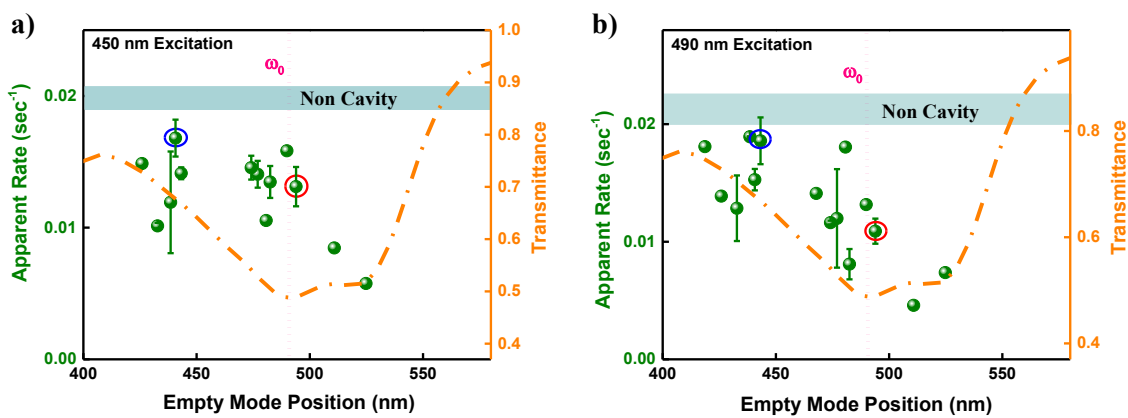

**Figure 4.** Kinetic action spectra of azopyrrole forward switching inside the cavity with varying mode position while pumping at a) 450 nm and b) 490 nm excitations, respectively. Non-cavity apparent rates are shown by the shaded green region. The transmission line shape of the azopyrrole molecule is given as a guide to the eye. Red and blue-circled data points represent the ON- and OFF-resonance reaction rates, which are shown in **Figure 3**. Each cavity is prepared independently, and the standard errors are calculated for those points in which the empty mode overlaps within 3 nm of λ/2 cavity mode position.

Tuning experiments are done by changing the thickness of the cavity, and, in turn, cavity mode position and reaction kinetics are plotted versus the empty cavity mode position (calculated from TMM), as shown in **Figure 4**. Several experiments were conducted in and around the absorption envelope, and it was found that the rate changes follow the same line shape as the absorber molecules. Excitation at 490 nm gives the kinetic action spectra of the forward photoisomerization process, and when the system is at ON-resonance, the reaction rate decreases drastically. The non-cavity rate constant is estimated to be $2.09 \times 10^{-2} \pm 0.09 \times 10^{-2}$ s$^{-1}$, and the cavity effect is getting muffled at OFF-resonance conditions. Photoexcitation at 450 nm also gives a similar signature (**Figure 4a**). Here, the FWHM was estimated between 455 and 545 nm. Altogether, tuning the cavity mode under ultra-strong coupling conditions shows an overall decrease in photoisomerization kinetics when exciting at the UP and $\omega_0$. In contrast, the LP state behaves exactly the opposite.

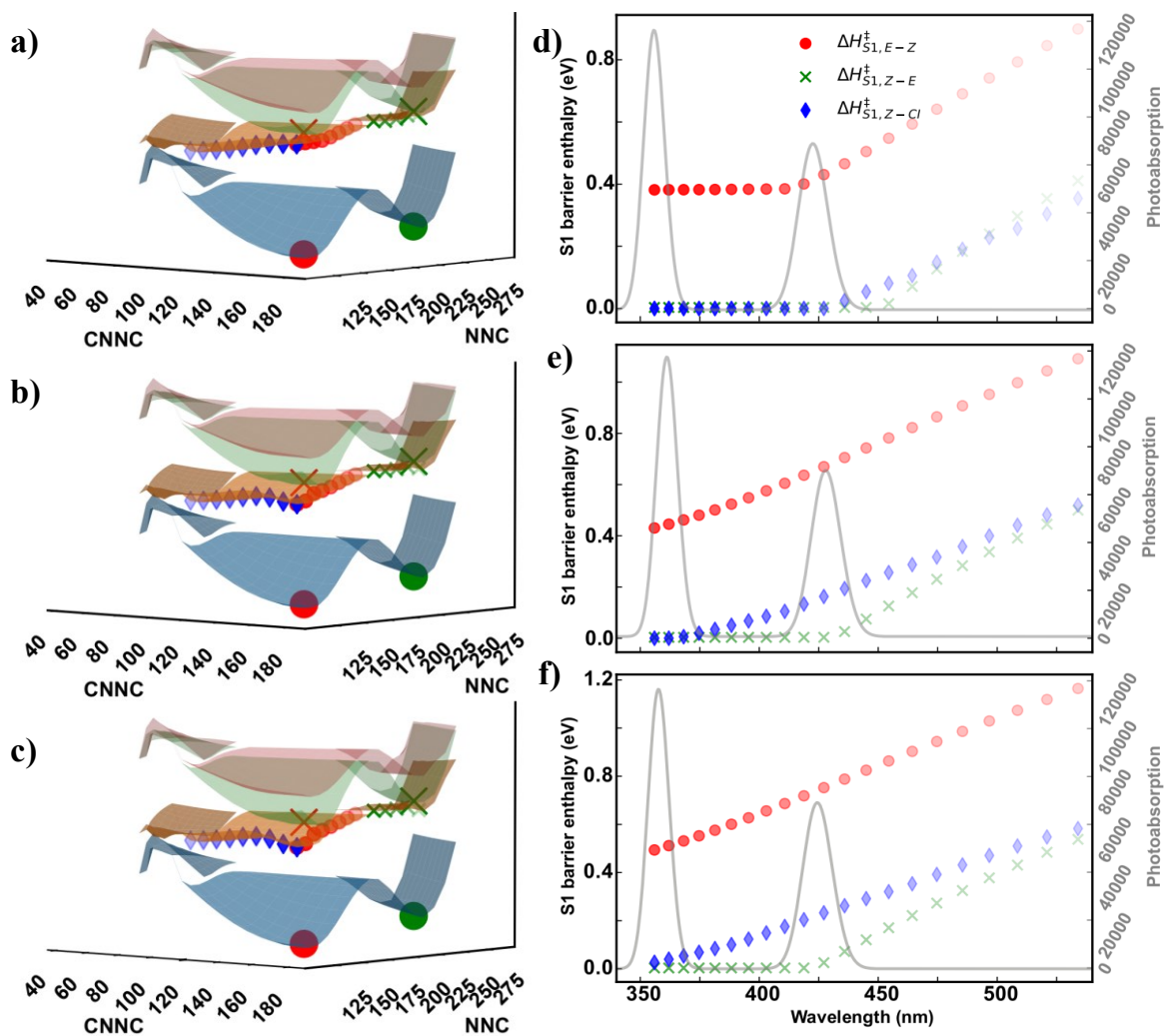

**Figure 5.** *Polaritonic modification of photoisomerization - theory:* a-c) represent modified polaritonic surfaces for $\{g/\omega_R =$ a) 0.0543, b) 0.1086, c) 0.1257$\}$ that result in the experimentally observed Rabi splitting ranging from weak (a; 1 w% azopyrrole at PSS), over strong (b; 1 w% azopyrrole), to ultra-strong (c; 1.5w% azopyrrole) for a frequency of the cavity mode chosen to be in resonance with the theoretically predicted absorption energy $S_0$ to $S_1$ of 427 nm. d-f) Transition-state barriers for different reaction paths on the $S_1$ (lower polariton) surface of the combined system with the absorption spectrum of the bare system as a grey line in the background. Higher transparency of a data point corresponds to a smaller matter contribution, i.e., a higher photonic fraction of the polariton.

Let us refine our understanding by taking a closer look at potential rate control via cavity-modified potential energy surfaces. We perform time-dependent density-functional theory (TDDFT) calculations with the ORCA5.0 quantum chemical code[33] by scanning C-N=N-C dihedral and N=N-C angle with constrained geometry optimization using the CAM-B3LYP functional[34] and including the toluene solvent implicitly[35] and a def2-TZVPD basis. Based on those geometries, we then calculate the ground and excited state potential energy landscape (PES). The absorption spectrum of the *E* isomer is shown as grey lines in **Figure 5** with a sizeable blueshift compared to the experimental measurements. Collective coupling between the ground and the *i*'th excited state is mediated by the coupling elements

$$g = \frac{1}{2}\overline{\mu_i} * E_{cav} * (N_E^{CNNC} * N_E^{NNC} + N_Z^{CNNC} * N_Z^{NNC}) \quad (3)$$

where $\overline{\mu_i}$ is the angular average of the transition dipole moment to the i'th excited state and $N_{E/Z}^{(C)NNC} = e^{-(R_{(C)NNC} - R_{E/Z})^2/\sigma^2}/\sqrt{2\pi\sigma^2}$ is a Gaussian distribution represented in the (C)N=N-C angles around the *E* (red dot and cross) and *Z* (green dot and cross) isomer with standard deviation in the respective angles of σ = 20 degree. This is a simplified model for the expected number of molecules participating in the collective interaction for a given configuration, i.e., the further away from *E* and *Z* isomers, the fewer molecules can contribute to the cavity PES (cPES) and the weaker its impact on the isomerization process. Please note that dark-state surfaces are not considered in this model. A matrix including 9 excited states plus an idealized loss-less cavity mode is then diagonalized to obtain the cPES[36] shown on the right-hand side of **Figure 5**. Isomerization of the *E* isomer into the Z isomer can take different paths, which can be characterized by either: (i) performing a CNNC inversion, e.g. using excess energy to overcome the $S_1$ barrier (red dots on $S_1$ in **Figure 5a-f**), or (ii) a barrier-less CNNC dihedral rotation (blue diamonds on $S_1$ in **Figure 5a-f**) via the $S_1$-$S_0$ CI from where relaxation to *E* and *Z* isomer is possible. As *E* and *Z*-isomer absorb the pump energy, the reverse process from *Z* to *E*-isomer, e.g., following pathway (i), results in the convergence towards PSS under constant illumination.

Longer wavelengths, i.e., large red-detuning, will formally increase the splitting between $S_2$ and $S_1$ at the loss of hybridization between light and matter. In other words, the $S_1$ surface continues to red-shift but turns into an effective photon replica of the ground-state $S_0$ surface without notable light-matter hybridization at the isomer. Hybridization appears now further away from the *E/Z* isomers, practically enhancing the $S_0$ "pocket" in which optically excited molecules could be trapped, ultimately reducing the isomerization rate. This essentially linear increase in cavity-induced barriers for increasing wavelengths (**Figure 5d-f**) is consistent with the experimental action spectra shown in **Figure 4**.

**Discussion:**

To get clarity on the overall data, the photoisomerization apparent rate per power density of the irradiation source is calculated for both concentrations (**Table 1**). Here, the ON-resonance cavity refers to the average of all the cavities that lie within the FWHM of the molecular absorption band. In the case of a strongly coupled system, the rate increases both at UP and 490 nm excitations. Whereas, in the case of LP excitation, only an initial increase can be observed within the strong coupling regime. Further, upon entering the weak coupling regime, the system goes back to the non-cavity rate. Non-cavity data somewhat follows the absorption profile as a maximum rate per power density is obtained at 490 nm excitation. Here, the LP state lies close to 550 nm, where the extinction coefficient is quite low; thus, less photoswitching is observed in a non-cavity.

**Table 1:** Estimated photoisomerization apparent rate per power density of excitation of cavity and non-cavity samples. *Please note that photoirradiation is always done from the bottom of the substrate.*

|  | Apparent rate/ Power Density (s$^{-1}$ W$^{-1}$ cm$^2$) | | | | | |
|---|---|---|---|---|---|---|
| **Strong coupling** | **450 nm** | | **490 nm** | | $^b$**LP excitation** | |
| Non-cavity | 20.95 ± 2.02 | | 34.69 ± 3.49 | | 14.42 ± 2.15 | |
| ON-resonance cavity | 50.05 ± 10.77 | $^a$87.82 ± 11.99 | 47.66 ± 7.61 | $^a$93.41 ± 12.46 | 25.36 ± 3.66 | $^a$11.03 ± 1.47 |
| **Ultra-strong coupling** | **450 nm** | | **490 nm** | | **LP excitation** | |
| Non-cavity | 33.31 ± 1.33 | | 35.67 ± 1.65 | | 8.78 | |
| ON-resonance cavity | 22.19 ± 5.82 | | 18.51 ± 6.21 | | 21.18 ± 0.37 | |
| OFF-resonance cavity | 22.62 ± 5.06 | | 28.01 ± 4.93 | | 20.37 ± 2.04 | |

$^a$two slopes are observed when the system moves from strong to weak coupling.
$^b$LP excitation position changes with cavity coupling conditions.

Next, we compared the apparent rate average for the USC regime for all the above excitation conditions. Here, both UP and 490 nm excitations show a decrease in the reaction rate. This decrement is more prominent if the ON-resonance cavity is at 490 nm. However, LP excitation is exactly the opposite, and the apparent rate per power density is ~2.5 times more than the non-cavity. Comparing the rate per power density for non-cavity at LP excitation, the values are lower in the case of USC conditions because here, the LP state is more red-shifted (>560 nm), where the transition probability of the uncoupled molecule is almost nil.

As discussed earlier, the dark-state overlap serves as a remarkably consistent figure of merit for the observed rate changes (**Figures 2 and 3**). Not only for the transition between strong and weak coupling, but even for the relative size of the rates between ultra-strong and strong coupling. This suggests that dark-state mediated relaxation processes dominate over cavity-induced modifications of the PES, at least near the transition from strong to weak coupling. At USC, however, cavity-modified potential-energy surfaces suggest a strong rate inhibition by trapping molecules in a dip of the S$_1$ surface near the *E/Z* isomers (**Figure 5**). The associated enthalpic barriers increase almost monotonously (in the investigated domain) with increasing wavelengths, which is consistent with the experimental action spectrum (**Figure 4**). Therefore, cPES can be an indicator for explaining rate modification in the USC domain.

**Conclusion:**

Electronic strong coupling can control photoisomerization processes as it involves an excited state process. Normally, photoconversion rates depend on the concentration of isomers and the photon flux. Here, by varying the collective coupling strength of the *E* and *Z*-isomers of the azopyrrole cooperatively, we have demonstrated the importance of strong coupling in funnelling optical absorption via the DS population in a photoisomerization reaction. Many factors control the kinetic rates, including the PES, the DS, and the excitation overlap of the collectively coupled system. In the present study, each of the parameters is estimated and compared. However, DS overlap seems to play a major role in this reaction under *collective* strong coupling. A very interesting observation is a complete rate modification upon strong-to-weak coupling transition in an on-the-go reaction. DS overlap provides an excellent figure of merit for those transitions, which is otherwise not evident from cavity-modified PES. We mapped excitation dependence reaction rates at both bright states and DS and clearly correlated them with available field distribution and overlap functions. In short, the optical strength of a macroscopic number of molecules exchanging energy coherently via the cavity mode acts as an antenna for any incoming optical drive. Excitation energy is then funnelled from the polaritonic states into the DS manifold (or grey states), where the isomerization proceeds as usual. The spectral overlap between polaritonic states and dark/grey states plays, conceptually, an equivalent role to the strong resonance dependence of Förster resonant energy transfer but is now macroscopically distributed over the entire coherence length of the cavity mode. Our observation emphasizes the role of DS spectral overlap, besides their visual absence, remains an integral part of cavity photochemistry. Surely, this conclusion is limited to the *collective* strong coupling regime. Additional measurements and theoretical investigations in the USC domain highlight that cavity-induced modifications of the PES remain a possible explanation under more extreme conditions. Further studies will be necessary to draw a comprehensive picture of the relative importance of DS-mediated relaxation and modified reactivity via cPES. In this work, we have identified the dynamic transition between two coupling domains as a potentially critical tool in the understanding of the fundamental mechanisms behind cavity-modified reactivity – preparing the ground for a leap to building a comprehensive understanding of QED chemistry.

**AUTHOR INFORMATION**

**Corresponding Author**

Jino George – Department of Chemical Sciences, Indian Institute of Science Education and Research (IISER), Mohali, Punjab 140306, India;

Email: jgeorge@iisermohali.ac.in; orcid.org/0000-0002-3558-6553

Christian Schäfer - Condensed Matter and Materials Theory, Department of Physics, Chalmers University of Technology, 412 96 Gothenburg, Sweden.
Email: christian.schaefer.physics@gmail.com

**Authors**

Pallavi Garg – Department of Chemical Sciences, Indian Institute of Science Education and Research (IISER), Mohali, Punjab 140306, India;

Jaibir Singh – Department of Chemical Sciences, Indian Institute of Science Education and Research (IISER), Mohali, Punjab 140306, India;

Ankit Kumar Gaur – Department of Chemical Sciences, Indian Institute of Science Education and Research (IISER), Mohali, Punjab 140306, India;

Sugumar Venkataramani – Department of Chemical Sciences, Indian Institute of Science Education and Research (IISER), Mohali, Punjab 140306, India;

**Funding Sources**

J. G. thank the SERB-core research grant (**CRG/2023/001122**), S. V. thank SERB-core research grant (**CRG/2023/003861**), and C. S. thank the European Union under the Marie Skłodowska-Curie Grant Agreement **No. 101065117**.

**Notes**

The authors declare no competing financial interest.

**ACKNOWLEDGMENT:**

P. G. thank UGC-CSIR, J. S. thank SPM-CSIR, A. K. G thank IISER Mohali for the PhD fellowships. J.G. thank DCS, and IISER Mohali for the use of the infrastructure facility.